# Variations in Tracking in Relation to Geographic Location

Nathaniel Fruchter, Hsin Miao, Scott Stevenson
Carnegie Mellon University
{nhf,hsinm,sbsteven}@andrew.cmu.edu

Rebecca Balebako
RAND Corporation
balebako@rand.org

*Abstract*—Different countries have different privacy regulatory models. These models impact the perspectives and laws surrounding internet privacy. However, little is known about how effective the regulatory models are when it comes to limiting online tracking and advertising activity. In this paper, we propose a method for investigating tracking behavior by analyzing cookies and HTTP requests from browsing sessions originating in different countries. We collect browsing data from visits to top websites in various countries that utilize different regulatory models. We found that there are significant differences in tracking activity between different countries using several metrics. We also suggest various ways to extend this study which may yield a more complete representation of tracking from a global perspective.

## I. INTRODUCTION

PRIVACY laws have been enacted worldwide with the purpose of protecting internet users' private information. Privacy laws can be divided into four main models [1] that differ in the scope and enforcement of regulation around digital privacy. The scope may include protections for all digital data, as in the *comprehensive model*; this is the model that has been adopted in the European Union. Alternatively, the scope of regulation protecting privacy may be limited and only protect certain types of data, such as health or education data, or certain classes of people, such as minors. This *sectoral model* is most notably adopted in the United States. The third model is *co-regulatory*; in this case, the scope may be similar to the comprehensive model, but in this model the rules are enforced by industry, as opposed to the state. Countries like Australia have adopted this approach. Finally, some countries have *no* legal or regulatory protection for digital privacy—a variety of locations fall under this regime, including the People's Republic of China [1], [2]. These models impact how countries handle privacy both legally and culturally, specifically in the realms of online tracking and privacy legislation.

Web tracking is implemented in a variety of ways, some of the most popular being third-party cookies and JavaScript tracking code. Commercial websites utilize a diverse plethora of trackers for various purposes such as targeted advertisements. Although privacy laws vary in different countries, there is currently a lack of information as to whether the number and types of trackers differ between countries, and whether this is impacted by different privacy regulation models. The purpose of this paper is to establish a empirical method for determining relationship between the amount of tracking and various countries that employ different privacy regulatory models.

In this study, we compared the amount of trackers on websites that operate in various countries with different privacy models. This paper offers three main contributions.

- We build, test, and describe an empirical, automated method for measuring the amount of web tracking in different countries, which can help determine the effectiveness of different privacy regulatory models.
- We examine the level of web tracking in four different countries representing three regulatory models, finding significant differences between countries.
- We investigate whether the location of the user or the location of the site is the factor that leads to differences in tracking between countries, finding that the site's country is more important than the visitor's country.

We have chosen Germany to represent the comprehensive model, the United States and Japan to represent the sectoral model, and Australia to represent the co-regulatory model. The sites that we are interested in are Alexa Top 250 sites [3] that have domains in multiple countries. We utilized Amazon Web Services to visit and crawl the data from the websites by servers in those countries.

We locate and identify these trackers using 3rd party HTTP requests and cookies. In addition, we identify ads from the websites by using a list provided by AdBlock browser extension [4]. Automation of the process is handled using the OpenWPM [5] tool which allows for synchronization across browsers and virtual machines ensuring that requests will occur at the same time.

In the following sections, we will first review some related work. Detailed descriptions of our method and experimental results are stated in Section III and IV. Discussion and possibilities for future work are described in Section V and VI.

## II. RELATED WORK

Privacy in the news seems inescapable; a general concern regarding the intrusiveness and pervasiveness of online tracking, advertising, and monitoring has caught the public attention. For example, concerns over the activities of social networking sites and advertisers such as Facebook bring up issues of anonymity and tracking in daily life [6]. Similarly, the level of privacy protection put into place by industry giants such as Google has come under scrutiny as jurisdictions with more comprehensive privacy regulations have called the effectiveness of their protections into question [7].

These worries also demonstrate the large amount of change that the Internet has undergone in a relatively short amount of time. As Mayer and Mitchell note [8], individual instances of web content have evolved from a single-origin affair into a conglomeration of "myriad unrelated 'third-party' websites," each facilitating anything from advertising to social media. This has been demonstrated by Krishnamurthy and Wills [9] in their longitudinal study, demonstrating what they term an "increasing aggregation of user-related data by a steadily decreasing number of entities." Furthermore, this explosion of third parties has existed an environment with little to no regulation until very recently [8], with advances only occurring in the comprehensive regulatory environment provided by the European Union.

We give a brief introduction to the privacy regulatory models. We then describe previous work in privacy-related web measurement. Finally, we provide background on best practices for web measurement methods.

### A. Privacy Regulatory Models

Privacy regulations differ around the world [10], [11]. The different regulatory models employed can be divided in several ways. While we use the taxonomy of regulatory models described below, other work has provided a more granular categorization of regulatory models [11], [12]. The empirical method described in this paper can be applied to either taxonomy of privacy regulatory models, as long as countries from all regulatory models are represented.

Different privacy regulation models around the world may have different impacts on the market, technology, and law [1]. In this work, we examine four models of privacy regulation.

1) *Comprehensive* regulatory models view privacy as a fundamental human right. They require companies and organizations protect personal information by placing limits on collection, use, and disclosure. A privacy authority agency enforces privacy laws. This comprehensive model is adopted in the European Union [1].
2) In a *sectoral* model, the government enacts privacy laws about a particular industry sector, for example in health or finance, but does not provide fundamental protection on privacy. The sectoral model is adopted in the United States [2].
3) A *co-regulatory* model relies on industries to develop their privacy policies for data protection. This is adopted in Australia.
4) Finally, a *mixed/no-policy* model is describes the regimes in which either privacy is not protected, or uses a mix of the other three policies. According to Swire and Ahmad, this model is adopted in the People's Republic of China [1].

### B. Measuring Advertising and Tracking Activity

Online behavioral advertising and tracking is a well-studied problem in the realm of privacy. While most users do perceive the potential benefits of this sort of targeting, many also express a "general, abstract notion of privacy violation" stemming from these advertising and tracking techniques [13]. Similar work concludes that these tracking techniques "[violate] consumer expectations" and notes that many users are generally unaware of the implications of behavioral advertising and tracking mechanisms such as cookies. These "misconceptions" may lead to an inability for users to make informed privacy decisions [14].

With this in mind, we decided to operationalize the presence of online tracking in two different ways: the presence of third-party cookies and the presence of third-party HTTP requests. Third-party cookies are already an established metric for the presence of advertising and tracking activity due to the fact they come from a source other the website a user is visiting. This usually indicates the presence of external content, something that often takes the form of an advertisement or tracking tool [8]. Testing for the presence of third-party HTTP requests is a natural extension of the cookie metric. As an HTTP request is made every time an external resource is loaded by the users web browser, analyzing requests unrelated to the main site the browser is accessing can help detect the presence of what Mayer and Mitchell term "third-party services" [8]. This also motivates our use of ad blocking lists as a heuristic (expanded upon in our discussion of methods) as these tools operate on a very similar detection principle.

### C. Privacy-related Web Measurement

In order to address and understand the impact of new web technologies on privacy, many efforts have been made to advance the field of privacy-related web measurement in recent years. Engelhardt et al. [15] have identified 32 studies that they categorize as "web privacy measurement studies." This category of study has great breadth, ranging from technical analyses of information leaked by web scripting languages [16] to empirical analyses of search engine personalization [17]. In this vein, numerous comparison-style studies have also been run, touching on diverse subjects such as discrimination in online advertising [18] and the effectiveness of online privacy tools [19].

The above studies make valuable contributions by taking on tasks like revealing the sources of potential privacy harms, detailing the effects of these third party entities, and taking a user-centric view to studying and enhancing privacy. However, they generally do not explore the impact of industry and country-level policy on the overall incidence of these third parties. Connolly comes the closest, performing an evaluation of various websites' compliance with the European Union's "Safe Harbor" privacy policy. Finding an astoundingly small subset of companies in compliance with Safe Harbor directives, Connolly discusses the "significant" privacy risk to consumers resulting from noncompliance [20]. Issues like these raise the necessity for a more comprehensive measurement of jurisdictional differences in tracking and advertising activity.

*D. Web Measurement Methodology*

Englehardt et al. conducted a study that reviewed general experimental frameworks and performed methodological analyses of extant web measurement studies. They found that web measurement studies are considered challenging for two reasons: causality and automation [15]. Controlled and randomized experiments are difficult in the dynamic, ever-changing web ecosystem [21]. Automation is difficult for several reasons, including that an automated script cannot always mimic real user behavior in browsers [15]. These difficulties has lead to some inconsistency and reinvention in web measurement. In order to address these issues, Engelhardt et al. authors developed a platform, OpenWPM [5], that addressed many of the issues of flexibility and scalability surrounding past web measurement studies. OpenWPM is a Python-based web-crawler framework using Selenium [22]. Due to its flexibility and convenience, it has been validated in several studies [5] [15]. This framework is utilized in the work in order to avoid further problems, especially those surrounding replication of effort and methodological inconsistencies.

## III. METHOD

We developed an automated method for measuring web privacy in different regulatory environments. We examined the quantity and type of web cookies set and third-party HTTP requests made when browsing to popular sites from different countries. To do so, we automated web browsing to the 250 sites (as determined by Alexa) in different countries at simultaneously. We collected data about the browsing sessions, including the HTTP requests and cookies. We then used a heuristic to examine results specifically related to web tracking. Finally, we used statistical analysis to compare the differences between countries.

For this method, we needed to overcome several technological hurdles, such as automating browsing from several countries in a controlled manner. To control for the impact of timing, we run the tests from multiple countries at the simultaneously [21]. Furthermore, we needed a method to determine whether the HTTP requests and cookies were third-party URLs and related to web tracking. In the next subsections, we describe how our method addresses these issues.

*A. Sourcing requests from several different countries*

We have chosen Germany to represent the comprehensive model, the United States and Japan to represent the sectoral model, and Australia to represent the co-regulatory model. Therefore, we sourced our data collection from four different locations.

To source an internet connection point at these various locations around the world, we used Amazon Web Services, or AWS [1]. AWS provides cloud-based virtual machines that can be configured in numerous ways. We installed OpenWPM on these machines and ran our tests from the cloud without having to rely on a proxy to set our location. AWS offers virtual machines in any of the following locations: Virginia (US), Ireland (EU), Frankfurt (EU), Oregon (US), California (US), Singapore (Asia), Sydney (AU), Sao Paolo (South America), and Tokyo (JP) [23]. This covers almost all of the regions we would like to examine – the only regions not represented are Russia and China which are currently not options when using AWS EC2. AWS employs a 'pay-for-what-you-use' model, so it is economically convenient to use. For example, running our study cost under $5 USD given Amazon's pricing schedule as of November 2014[2].

*B. Selecting which sites to visit*

For each country, we crawled the top 250 sites for that country using the Alexa list by country [3]. Typically these are top level domains and not subpages within a site. While there was some overlap of sites between countries, such as `google.com` and wikipedia.org, there were differences between the country lists. First, some domains were specific to the country, such as `facebook.de` in Germany. Second, many sites were specific to that country or language. One example of a website specific to Germany is the domain for a popular news journal Der Spiegel (`spiegel.de`), which was not seen on the other country lists.

*C. Automating the web crawls*

Our next step was to automate the data collection. We collected a number of metrics related to tracking, including the number of cookies and HTTP requests. Engelhardt et al.'s OpenWPM platform is a purpose-built web measurement platform that logs a large amount of web session data in a standardized SQLite database format, making it the perfect tool for our study. We utilized the most recent publicly available version of OpenWPM, 0.2.0, for the data collection portion of our study and used the platform's API to programmatically crawl a list of the top 250 websites as defined by Alexa [3]. OpenWPM's Firefox backend was used for the crawl with both JavaScript and Flash enabled.

Two variables of interest are located within different SQLite tables generated by OpenWPM with each crawl: `cookies` and `http_requests`. We extracted the domains of cookies and the URLs of HTTP requests from these two tables by using Python's `sqlite3` library module.

*D. Extracting Third-party HTTP requests and cookies*

In this study, we were not interested in analyzing first-party cookies and HTTP requests, as these are often not considered privacy invasive [24]. Therefore, we had to extract the third-party elements of our collected data. In order to further analyze third-party cookies and HTTP requests, we set a rule to determine whether the URL in a record is related to the website where the record was extracted. To be more specific, if the URL in a record does not contain the domain name of the website we are currently visiting, it is a third-party cookie or HTTP request. For example, if a cookie is extracted from `amazon.de` and the URL is `fls-eu.amazon.de`, it is a first-party cookie because the base domain is identical. In contrast, if a cookie also extracted from `amazon.de` has the domain `zanox.com`, then the domains are not identical and it is a third-party cookie. By implementing these procedures, we can use statistical tools to analyze the collected data.

---

[1] http://aws.amazon.com
[2] We used `t2.small` instances running Linux at $0.026 per CPU hour.
[3] http://www.alexa.com/topsites/

*E. Tracker Heuristic: AdBlock "easylists"*

Not all the URLs identified using the above method are necessarily related to advertising or web tracking. They may also be first-party content hosted on content management networks or separate servers maintained by the first party. Therefore, we used an additional heuristic to determine which URLs were related to web tracking and advertising.

AdBlock Plus [4] is a popular browser extension available for both Firefox and Chrome which allows users to filter and block elements on a webpage according to user-specified rules. As evidenced by the extension name, this capability is most often used in service of blocking advertisements, tracking code, or other content deemed annoying, invasive, or objectionable. Due to its open source nature and large, international user base, AdBlock Plus provides a unique resource: a large, crowd-sourced list of rules that allows us to detect the presence of advertising or tracking assets within a list of URLs and page elements. These rules are compiled in two "easylists" [25] provided on the AdBlock website, with one focused on ad-blocking rules and the other focused on tracker-blocking rules.

Using a similar approach to the one detailed in the last section, we extracted the full URLs of HTTP requests and responses from the OpenWPM crawl database using Python and the sqlite3 library. We then used the `adblockparser` [26] Python module to match the extracted HTTP request and response URLs against the two sets of AdBlock rules mentioned above. The number of positive ad or tracker hits were aggregated by domain, country, and rule set in order to produce summary statistics for use in further analysis. (In this paper, we are using the term "hit" to denote one of these positive pattern matches between an AdBlock easylist rule and the domain or URL seen in an HTTP request or cookie.)

## IV. RESULTS

We ran our script on the top 250 sites for each of our four countries. We collected all the HTTP requests and cookies from these browsing sessions and then used the heuristic and algorithm described in the previous section to identify probable tracker activity. We found that visiting the sites from the US yielded the most third-party HTTP requests and third-party cookies. Through additional comparisons with a dataset based on the top 500 sites globally, we also found indications that other factors besides user origin may come into play. For example, a website's country of origin or a server's physical location may have even more impact than a user's geographic location. Differences between countries with the same regulatory model could also indicate differences stemming from other cultural, business, and political factors.

TABLE I
RANK OF THE NUMBER OF THIRD-PARTY HTTP DOMAIN REQUESTS AMONG DIFFERENT COUNTRIES USING A KRUSKAL-WALLIS TEST. THE US HAD SIGNIFICANTLY MORE THIRD-PARTY HTTP REQUESTS THAN THE OTHER COUNTRIES.

| country | Rank |
|---------|--------|
| US | 575.00 |
| AU | 511.79 |
| DE | 492.52 |
| JP | 406.69 |

*A. Evaluation Metric: Third-Party Cookies and Requests*

The goal of our study was to discover the variation in tracking activity between different countries. Due to the categorical-quantitative nature of our data, a one-way ANOVA model was deemed appropriate for our analyses. More specifically, all data was analyzed using the nonparametric Kruskal-Wallis test due to the the variable sample size and non-normality of our samples. This allowed comparisons of tracking activity across varying levels of country, our independent variable.

There are some dependent variables we used for further analysis. First, we analyzed the number of third-party cookies and HTTP requests, which is closely related to online tracking activity. Second, we examined first and third-party cookies and HTTP requests to see whether the ratios were identical in different countries. Since the raw counts of cookies and requests were not normalized, this helped us to understand the relative proportions of the two types of cookies and requests for different segments of our data. Additionally, the number of first-party cookies and HTTP requests were analyzed because some sites (e.g., Google) are both an analytics provider and a service provider, as such they may use other methods besides third-party cookies to track users.

*B. Third-party HTTP requests*

We compared the number of third-party HTTP domain requests among different countries. Table I shows the average rank for each country in Kruskal-Wallis test. We found that the difference of the numbers of third-party domain of HTTP requests among our four countries are significant ($\chi^2 = 43.863; df = 3; p < 0.0005$). We also found that there are more third-party HTTP requests in the US compared to Germany and Australia ($\chi^2 = 10.752; df = 1; p = 0.001$). The differences between Germany and Australia were not significant. Moreover, there were more third-party HTTP requests in Germany and Australia compared to Japan ($\chi^2 = 39.709; df = 1; p < 0.0005$).

*C. Cookies*

We also compared the number of third-party and first party cookies among different countries. Table II shows the average rank of number of third-party cookies for each country in Kruskal-Wallis test. Although the difference in total number of first-party cookies is not significant, the difference of number of third-party cookies is significant ($\chi^2 = 13.147; df = 2; p = 0.004$). This implies that, generally, a visitor to one of the top 250 sites from the United States would be exposed to a comparatively greater amount of web tracking code.

TABLE II
RANK OF THE NUMBER OF THIRD-PARTY COOKIES AMONG DIFFERENT
COUNTRIES USING A KRUSKAL-WALLIS TEST.

| country | Rank |
|---|---|
| US | 499.14 |
| DE | 445.51 |
| AU | 438.53 |
| JP | 411.91 |

TABLE III
CORRELATION BETWEEN NUMBER OF HTTP REQUESTS AND COOKIES

| Country | r |
|---|---|
| AU | 0.691 |
| DE | 0.634 |
| JP | 0.778 |
| US | 0.715 |

We found similar results when comparing the number of domains in third-party HTTP requests. There are more third-party cookies in the US compared to Germany ($\chi^2 = 4.111; df = 1; p = 0.043$) and Australia. Also, the difference between Germany, Australia, and Japan is not significant.

### D. Correlation between HTTP requests and cookies

Table III shows the correlation between the number of third-party domain for HTTP requests and third-party cookies. We found that in these countries these two variables are strongly correlated. Furthermore, the high correlation coefficients for all countries in our data demonstrates a good level of internal validity for our two measures of online tracking. This indicates a sufficient level of confidence in the underlying construct that we are attempting to measure. Additionally, this also shows that HTTP requests can be a good measure of tracking in addition to the already established cookie measure.

### E. Evaluation Metric: AdBlock rules

*1) Origin-dependent tracking activity:* One crucial phenomenon to test for is the presence of origin-dependent tracking activity–in other words, we wanted to determine if the origin of the user was important or if the origin of the website was important. For example: if user A visits example.com from country A and user B also visits example.com at the same time, but from country B, will they receive the same type and number of trackers? This analysis was done to determine how heavily geographic factors need to be controlled for in this (and other) studies. Finding that the users' source matters may indicate interesting, adaptive behavior by tracking companies that could warrant further investigation.

To this end, we crawled Alexa's list of the top 500 global sites from all four of our server locations at identical times and compared matches against Adblock's tracking EasyList. Controlling for outliers, nonparametric tests of both the absolute number of hits by country and the proportions of hits by country show no significant difference (n hits: $\chi^2 = 0.805; df = 3; p > 0.84$, proportion: $\chi^2 = 0.172; df = 3; p > 0.98$). Because of this, we can conclude that the impact of request origin will not be a significant factor for us within the scope of our experiment.

This conclusion is further bolstered by an interesting possibility that stems from a comparison of our country-specific datasets with our new, global dataset. Looking at the series of pairwise comparisons for the top 500 sites (see Table VII), none of the differences between countries are significant (all $p > 0.71$). This indicates that it may be the website's country origin, not the user's, that matters in terms of tracking activity present. However, there may be other factors that account for this difference in variation, something that will be expanded on in our discussion.

*2) More trackers than ads:* There were significant differences in type of hit (trackers vs. advertisements) within the same top 500 sites. The proportion of requests associated with trackers was significantly higher than the proportion associated with advertisements ($\chi^2 = 45.1; p < 0.0001$). A pairwise comparison across the top 500 sites showed that trackers accounted for approximately 2% more requests than advertisements ($95\% CI[0.015, 0.021]$). This is significant considering the overall proportion of requests for both ads and trackers is 5.4% ($SEMean = 0.0009, 95\% CI[0.052, 0.056]$). Since trackers, as opposed to ads, do not usually have a visual element, this may imply a "tip of the iceberg" issue for users. While awareness of online advertising may be relatively high, the invisibility of online tracking for the typical user may lead to a false sense of privacy for some.

*3) Differences by country:* Based on our limited sample of countries per regulatory model, we do not draw conclusions about the regulatory models themselves. We do find interesting results when examining each individual country in a series of pairwise comparisons between the top 250 sites in each country. Differences in the proportion of total HTTP requests associated with trackers differs significantly and may imply the presence of significant variation beyond what can be explained on the country or model level.

*4) More tracking-related requests in the United States:* A pairwise examination of the proportion of HTTP requests related to tracking activity (operationalized as the proportion of requests that matched an Adblock rule) show that United States has significantly more tracking activity compared to all of our other countries. While the differences varied by country, each comparison showed a significantly greater (at least $p < 0.02$) percentage of tracking requests, ranging from less than 1% (US-AU) to more than 3% (US-JP). Table VI displays these pairwise tests, along with confidence intervals, in more detail.

*5) Differences within the sectoral model:* It is especially interesting to note the comparisons between our two sectoral model countries, the United States and Japan. Even though they ostensibly have the same regulatory model, the United States showed a significantly greater (all $p < 0.02$) amount of tracking-related HTTP requests (anywhere from 2.8% to 4% more). Considering the average number of requests per page is over 100, even a 4% increase in tracking-related requests could indicate the loading of 4 to 5 more tracking elements or scripts per browsing session. This can be seen in the summary statistics shown in Table IV, which displays the mean number of HTTP requests, hits, and hits as a proportion of total requests for the browsing sessions.

TABLE IV
SUMMARY STATISTICS FOR ALL TRACKING-RELATED HTTP REQUESTS

| N | Mean Requests (SD) | Mean Hits (SD) | Mean Proportion Hits (SD) |
|---|---|---|---|
| 1931 | 111 (116) | 6.54 (7.7) | 0.06 (0.05) |

TABLE V
SUMMARY STATISTICS BY COUNTRY FOR TRACKING-RELATED HTTP REQUESTS

| Country | Mean (Number_Requests) | Mean (Number_Hits) | Mean (Proportion_hits) | Std Dev (Number_Requests) | Std Dev (Number_Hits) | Std Dev (Proportion_hits) |
|---|---|---|---|---|---|---|
| AU | 99.19 | 6.83 | 0.06 | 80.70 | 7.0 | 0.05 |
| DE | 121.04 | 5.70 | 0.05 | 160.74 | 6.31 | 0.05 |
| JP | 103.15 | 4.10 | 0.05 | 101.64 | 4.82 | 0.05 |
| US | 120.59 | 9.34 | 0.08 | 105.10 | 10.41 | 0.05 |

TABLE VI
PAIRWISE COMPARISONS BETWEEN COUNTRIES FOR TRACKING HITS

| Country A | Country B | Z | p | 95% CI For Change |
|---|---|---|---|---|
| US | JP | 10.42 | <.0001 | [0.028, 0.040] |
| US | DE | 7.77 | <.0001 | [0.018, 0.031] |
| US | AU | 2.57 | <.02 | [0.001, 0.014] |
| JP | DE | -3.64 | <.0005 | [-0.013, -0.002] |
| DE | AU | -5.29 | <.0001 | [-0.021, -0.009] |
| AU | AU | -8.33 | <.0001 | [-0.031, -0.019] |

TABLE VII
PAIRWISE COMPARISONS BETWEEN COUNTRIES FOR TOP 500 GLOBAL SITES

| Country A | Country B | p |
|---|---|---|
| JP | DE | 0.855 |
| JP | AU | 0.963 |
| US | DE | 0.859 |
| DE | AU | 0.838 |
| US | JP | 0.739 |
| US | AU | 0.714 |

## V. DISCUSSION

### A. Outliers

We are also interested in outliers about tracking behaviors in the websites. For example, the US, nydailynews.com has the most number of third-party cookies in top 250 websites. There are 6,546 third-party cookies set when when that website is visited. Other news websites including foxnews.com, sfgate.com, drudgereport.com and nypost.com all have more than 900 third-party cookies. Similarly, we found that news websites also play important roles in Japan and Australia's third-party cookie statistics. The site theaustralian.com.au has 1,819 third-party cookies on its website, which is the third most in their top 250 websites. In addition, in Japan, reuters.com has 1,827 third-party cookies, which is the most in top 250 websites. The finding is interesting because it implies that news websites rely on third-party cookies heavily in the US, Japan, and Australia. However in Germany, the tracking behaviors are not similar to other three countries because most of third-party cookies are set by shopping websites instead of news websites.

### B. Other factors

Currently, we do not have enough data to conclusively say whether the different privacy regulatory models are actually statistically different from one another in practice. However, we did find evidence that privacy regulatory models alone may not indicate the level of technological privacy users get. For instance, we noticed that the US had many more tracking indicators than Japan overall, even though they both follow the sectoral model. We are unsure of exactly why this is the case but we suspect that it may be due to cultural differences or perhaps the types of websites that are popular. It could be the case that the popular sites in Japan fall under a particular sector that is more regulated than those in the US.

Another possibility along these lines is that tracking, advertising, and the sale of customer data is not the most popular business model for websites in Japan – another factor that could lead to the differences in tracking we saw. Furthermore, this type of motivation could actually be related to our findings in the realm of news websites. Due to the shifting media landscape in many countries, newspapers and other journalistic organizations are constantly looking for new sources of revenue. Some sources put online advertising at roughly 20 percent of advertising revenue and this, along with other cultural and corporate factors, may contribute to the disproportionately large amount of advertising and tracking found on news sites [27].

### C. Limitations

A study collecting data from sources as dynamic as a popular website may encounter several issues with external validity. For example, the tracking activity present on a site may not be entirely deterministic given a certain page load – factors such as time of day and previous user activity may lead to differing types of activity behind the scenes [28]. Further confounds such as the automated nature of our data gathering process may introduce other sources of variation; for example, many EU-targeted sites do not set cookies unless a user explicitly opts in as a result of regulatory action [29]. While some related confounds like time of day were controlled for, the numerous sources of variation may warrant a follow-up study to assess the external validity of our data collection methods. Finally, our we recognize that these factors lead to an imperfect operationalization of tracking. Indeed, the complex inter-session and inter-device nature of modern tracking technology provide challenges for the web measurement and privacy research community. However, these challenges do provide promising avenues for future work and investigation.

*D. Future work*

Since we were unable to access a node in China and Russia from which to run our script, we have no direct representation of the no privacy model regions. We initially thought that this would directly affect our ability to measure tracking properly but our results have shown that where we connect to a particular website from may have very little to do with tracking. This result is based on a small sample (just the US and Japan) so we would also like to verify this fact over a longer period of time and with more countries prior to making a concrete conclusion. China may be an exception to this finding since they have the "Great Firewall of China" in place which may distort our results.

Russia is another interesting case and doesn't have the complication of a national firewall. Russia's government has been taking an increasingly aggressive interest in the internet, recently going so far as commandeering the Vkontakte, the 'Facebook of Russia' [30]. Extending our study to incorporate these two countries seems very promising since it could yield results that are very different from what we have seen in our current study. In the case of China, AWS EC2 is currently in open beta (for Chinese residents only) for nodes in Beijing. Setting up a node there may be possible in the near future. This would also give us the ability to measure and compare tracking in China from inside and outside the firewall.

It may be valuable to conduct this study again in the future as well. Doing so would allow comparison of tracking throughout time and such time-series data could grant us greater causal and explanatory power in a variety of situations. For instance, there may be value in examining changes in tracking after privacy-related news or policy events. If Do Not Track becomes a widely accepted standard, how different will the tracking landscape look? Would tracking increase or decrease for people not utilizing Do Not Track? Similarly, we could examine if major regulatory rulings by relevant agencies create noticeable change. If the FTC steps up enforcement in a particular area, is there a visible effect of that ruling?

Another valuable extension to our study would be to more deeply examine other methods of tracking. Third-party cookies, third-party HTTP requests, and AdBlock rules don't tell the whole story. For example, even though Google has very few third-party cookies or requests, it is well known that Google tracks users throughout their systems using various internal metrics [31]. In a similar vein, many major service providers like Google are also their own analytics providers. We do not account for this possibility in our study, but developing methods for doing so may reveal a more complete picture. In a similar vein, tracking is not a web-only construct; similar "first-party" tracking metrics could look very different on mobile, or other novel platforms.

Finally, we would like to investigate and better understand more subtle country-level differences. For example, what is causing the US to have much more tracking than Japan even though they are both sectoral countries? We can hypothesize that this may be cultural, or that the popular websites in Japan differ in category from those in the US or that the popular sites in Japan may fall under the regulation of a stricter sector. Questions like these also leave open the possibility of exploring more complex issues at the intersection of privacy, policy, and culture. For instance, richer data at a more granular level could help highlight the interaction between the effects of culture on Internet use, businesses' data collection habits, and regulation.

Since none of these questions can be fully explored with our data, additional research would have to be done to confirm or deny these assumptions. We expect collaboration with economists and local legal and cultural experts could improve understanding in this area. However, once these questions have been answered satisfactorily, we may be closer to evaluating how effective the regulatory models actually are for protecting end users—certainly a question of great importance to all stakeholders. Such information would also open up many relevant paths of inquiry and discussion for the privacy and policy communities, from the challenges of defining a privacy standard to the difficulties of translating regulation into compliance and enforcement in a global, networked environment.

## VI. CONCLUSION

Going into this experiment, we assumed that there would be a significant difference in tracking between countries in different privacy regulatory models. We expected to see the most tracking in the no-model and sectoral model countries, less in the co-regulatory model, and even less in the comprehensive model. To examine this, we developed an empirical, repeatable method to evaluate web tracking in different countries. This method can be used in future studies to measure the impact of policy changes.

We were also interested in determining if the country the website is based in, versus the country we are connecting from, plays a role in the amount of tracking. Due to our limited sample size, though, we were not able to draw strong conclusions regarding the regulatory models themselves. However, we were able to quantify many interesting variations in tracking behavior between countries and provide several directions for relevant future work to further investigate these variations. We were able to conclude that there were significant differences in tracking activity between different countries using several metrics.

The subtle country level difference indicate that the privacy regulatory model are not the sole factor impacting web tracking. For example, the US has much more tracking than Japan even though they are both sectoral countries. This may be cultural, as the popular websites in one country may differ in category from those in another country due to a cultural preference to view different types of websites. Alternatively, the popular sites in a country may fall under the regulation of a stricter sector, or web sites (such as national news organizations) may enjoy financial support that allows them to rely less on advertising. These options highlight that privacy may be impacted by regulation and culture that is not directly about privacy or data sharing. Further exploration is needed on these complex issues at the intersection of digital privacy, policy, and culture.